\documentclass[twocolumn]{aastex631}

\usepackage{multirow} 
\newcommand\ions[2]{{#1}\,{\sc #2}} 

\accepted{November 05, 2023}

\submitjournal{AJ}

\shorttitle{Near-IR spectroscopy for wide binary component stars}
\shortauthors{Lim et al.}

\begin{document}

\title{Chemical homogeneity of wide binary system: An approach from Near-Infrared spectroscopy}

\correspondingauthor{Dongwook Lim}
\email{dwlim@yonsei.ac.kr}

\author[0000-0001-7277-7175]{Dongwook Lim}
\affiliation{Center for Galaxy Evolution Research \& Department of Astronomy, Yonsei University, Seoul 03722, Republic of Korea}
\affiliation{Zentrum f\"ur Astronomie der Universit\"at Heidelberg, Astronomisches Rechen-Institut, M\"onchhofstr. 12-14, 69120 Heidelberg, Germany}

\author[0000-0002-9859-4956]{Andreas J. Koch-Hansen}
\affiliation{Zentrum f\"ur Astronomie der Universit\"at Heidelberg, Astronomisches Rechen-Institut, M\"onchhofstr. 12-14, 69120 Heidelberg, Germany}

\author[0000-0003-4364-6744]{Seungsoo Hong}
\affiliation{Center for Galaxy Evolution Research \& Department of Astronomy, Yonsei University, Seoul 03722, Republic of Korea}

\author[0000-0002-6154-7558]{Sang-Hyun Chun}
\affiliation{Korea Astronomy and Space Science Institute, 776 Daedeokdae-ro, Yuseong-gu, Daejeon 34055, Republic of Korea}

\author[0000-0002-2210-1238]{Young-Wook Lee}
\affiliation{Center for Galaxy Evolution Research \& Department of Astronomy, Yonsei University, Seoul 03722, Republic of Korea}


\begin{abstract}
Wide binaries, with separations between two stars from a few AU to more than several thousand AU, are valuable objects for various research topics in Galactic astronomy. 
As the number of newly reported wide binaries continues to increase, studying the chemical abundances of their component stars becomes more important. 
We conducted high-resolution near-infrared (NIR) spectroscopy for six pairs of wide binary candidates using the Immersion Grating Infrared Spectrometer (IGRINS) at the Gemini-South telescope. 
One pair was excluded from the wide binary samples due to a significant difference in radial velocity between its component stars, while the remaining five pairs exhibited homogeneous properties in 3D motion and chemical composition among the pair stars. 
The differences in [Fe/H] ranged from 0.00 to 0.07~dex for these wide binary pairs.
The abundance differences between components are comparable to the previous results from optical spectroscopy for other samples. 
In addition, when combining our data with literature data, it appears that the variation of abundance differences increases in wide binaries with larger separations. 
However, the SVO2324 and SVO3206 showed minimal differences in most elements despite their large separation, supporting the concept of multiple formation mechanisms depending on each wide binary. 
This study is the first approach to the chemical properties of wide binaries based on NIR spectroscopy. 
Our results further highlight that NIR spectroscopy is an effective tool for stellar chemical studies based on equivalent measurements of chemical abundances from the two stars in each wide binary system.
\end{abstract}

\keywords{
Wide binary stars(1801) --- 
Chemical abundances(224) ---
Stellar abundances(1577) --- 
High resolution spectroscopy(2096) --- 
Near infrared astronomy(1093) 
}

\section{Introduction}\label{sec:intro}
The discovery of co-moving wide binary candidates in the Milky Way has significantly increased based on high-precision parallax and proper motion solutions for a vast number of stars provided by Gaia Data Release \citep{GaiaCollaboration2023}.
For instance, \citet{El-Badry2021} recently reported a million wide binary candidates from the Gaia data, which is a stark contrast to the 1147 candidates revealed two decades ago from the New Luyten Two-Tenths (NLTT) Catalog \citep{Chaname2004}.
As the number of reported wide binary systems increases, numerous studies have been undertaken not only to understand their origin and evolution but also to explore various topics in Galactic astronomy through wide binary samples.

The formation of wide binary systems has been a long-standing problem, as two component stars with large separation cannot be easily formed from a single collapsing cloud.
Consequently, various scenarios have been proposed, such as formation during the early dissolution phase of young star clusters \citep{Kouwenhoven2010, Moeckel2011}, dynamical unfolding of higher-order systems \citep{Reipurth2012, Elliott2016}, formation by turbulent fragmentation \citep{LeeJE2017}, from adjacent pre-stellar cores \citep{Tokovinin2017}, in tidal streams of stars and globular clusters \citep{Penarrubia2021}, and in the turbulent interstellar medium \citep{Xu2023}. 
All of these scenarios suggest a similar chemical composition for the components of wide binaries, as already reported in various studies \citep[e.g.,][]{Desidera2004, Desidera2006, Lim2021}.
However, narrowing down a specific formation scenario remains challenging, and it is likely that wide binaries have formed through multiple channels depending on their chemical and kinematic properties.
Furthermore, due to the low binding energy and similar chemical properties between component stars, wide binary systems serve as valuable tools for studying mass constraints of massive compact halo objects \citep[MACHOs;][]{Yoo2004, Quinn2009, Quinn2010, Tian2020}, the validity of the chemical tagging technique \citep{Hawkins2020}, calibration of spectroscopic survey data \citep{Buder2021, Niu2023}, properties of accreted dwarf galaxies \citep{Lim2021, Nissen2021}, the connection between field stars and cluster stars \citep{Gruner2023}, and star-to-planet interactions \citep{Saffe2017, Oh2018, Ryabchikova2022}.

Most of these studies rely on high-precision chemical abundance measurements of stars in each wide binary. 
Therefore, high-resolution spectroscopy for co-moving wide binary candidates is essential to identify whether they are coeval binaries or randomly co-moving pairs, as well as for diverse studies on the Milky Way.
Recently, extensive spectroscopic observations of Milky Way stars have been conducted through surveys and individual observations.
One noteworthy observation is high-resolution spectroscopy in the near-infrared (NIR) wavelength region.  
This observation offers several advantages compared to the optical spectroscopy, such as obtaining high-quality data for cool stars that are brighter in the NIR region and stars located in regions of high extinction, like the Galactic bulge. 
In addition, high-resolution NIR spectroscopy enables precise measurements of abundances of volatile elements and molecules, which are crucial for exoplanet atmosphere studies \citep[e.g.,][]{Line2021}. 
Several NIR spectrographs, including Warm Near infrared Echelle spectrograph to Realize Extreme Dispersion \citep[WINERED;][]{Ikeda2016} and Immersion Grating Infrared Spectrometer \citep[IGRINS;][]{Mace2018}, are now employed in various fields of stellar astronomy.
However, due to the small number of stellar chemical abundance studies in the NIR spectral range, it is necessary to examine efficient methods to determine atmosphere parameters and chemical abundances, and to assess their reliability \citep[e.g.,][]{Lim2022, Nandakumar2023}.
Although  wide binaries, which have been utilized to check the validity of spectroscopy, can be a good test-bed for stellar high-resolution NIR spectroscopy, such observations have not been made yet. 

In this study, we obtained the first high-resolution NIR spectroscopic data for several wide binary candidates using the IGRINS instrument. 
Our primary objectives were to investigate the general properties of wide binaries, particularly for those with larger separations, and to identify any peculiar wide binary systems based on their chemical and dynamical properties.
In addition, we aimed to assess the validity of NIR spectroscopy for stellar chemical abundance studies. 
This paper is organized as follows. 
In Section~\ref{sec:obs}, we provide details on target selection, observation, and data reduction processes. 
The procedures for determining atmosphere parameters and measuring chemical abundance are presented in Section~\ref{sec:spec}.
We provided the results of the chemical composition of the observed stars in Section~~\ref{sec:result}, and these results are discussed in Section~\ref{sec:trend}. 
Finally, Section~\ref{sec:concl} is a summary and conclusion of our study.

\section{Observations and data reduction }\label{sec:obs}
\subsection{Target selection}\label{sec:sub:target}
Our spectroscopic observations were conducted during two observing runs in the 2021B and 2023A semesters. 
For each observing run, we selected target wide binaries from two different catalogs, each serving different purposes.

The targets for the 2021B observing run were selected from the catalog of \citet{Jimenez-Esteban2019}, which reported 3,741 co-moving groups based on Gaia DR2 \citep{GaiaCollaboration2018}. 
After cross-matching this catalog with NIR photometric data from the 2MASS catalog \citep{Skrutskie2006}, we selected 28 bright co-moving pair candidates (8.0 $<$ Gaia~$G$ $<$ 12.0; 7.0 $<$ 2MASS~$K_{s}$ $<$ 10.0) with a separation between the component stars larger than 10,000~AU. 
These selection criteria were defined to investigate the chemical homogeneity of wide binaries with wider separation, as well as to validate the NIR spectroscopic technique. 

For the 2023A observing run, we focused on wide binaries suspected to have accretion origin \citep[see, e.g.,][]{Lim2021, Nissen2021}. 
To do this, we selected target candidates exhibiting large proper motion and observable conditions at the Gemini-South telescope (6.0 $<$ Gaia~$G$ $<$ 15.0; $\mu$ $>$ 50 mas~yr$^{-1}$) from the SUPERWIDE catalog of \citet{Hartman2020}.
This catalog reported 99,203 wide binary systems based on Gaia DR2, providing an advantage in tracing accreted wide binaries due to its inclusion of a higher number of co-moving pair candidates with large proper motions compared to the catalog of \citet{Jimenez-Esteban2019}.
After updating the astrometric solution with the Gaia EDR3 \citep{GaiaCollaboration2021} and incorporating NIR photometric data from 2MASS, we pre-estimated the orbital energy ($E$) and angular momentum ($L_{Z}$) of the co-moving stars with the Galactic potential of \citet{McMillan2017}, assuming a line-of-sight velocity of 0 km~s$^{-1}$. 
We note that while the line-of-sight velocities of many target stars have been recently updated by Gaia DR3 \citep{GaiaCollaboration2023}, this information was not available during the preparation of our observation.
Finally, we selected 8 co-moving pairs in the dynamical domains of various accretion events, such as Sequoia, Gaia-Enceladus, and Helmi-streams, referring to the criteria of \citet{Massari2019}.

However, due to the telescope schedule and unfavorable weather conditions, only 10 and 1 co-moving pairs were actually observed during the 2021B and 2023A semesters, respectively, and 5 pairs were analyzed in this study (see Section~\ref{sec:sub:obs} below). 

\begin{figure}
\centering
   \includegraphics[width=0.45\textwidth]{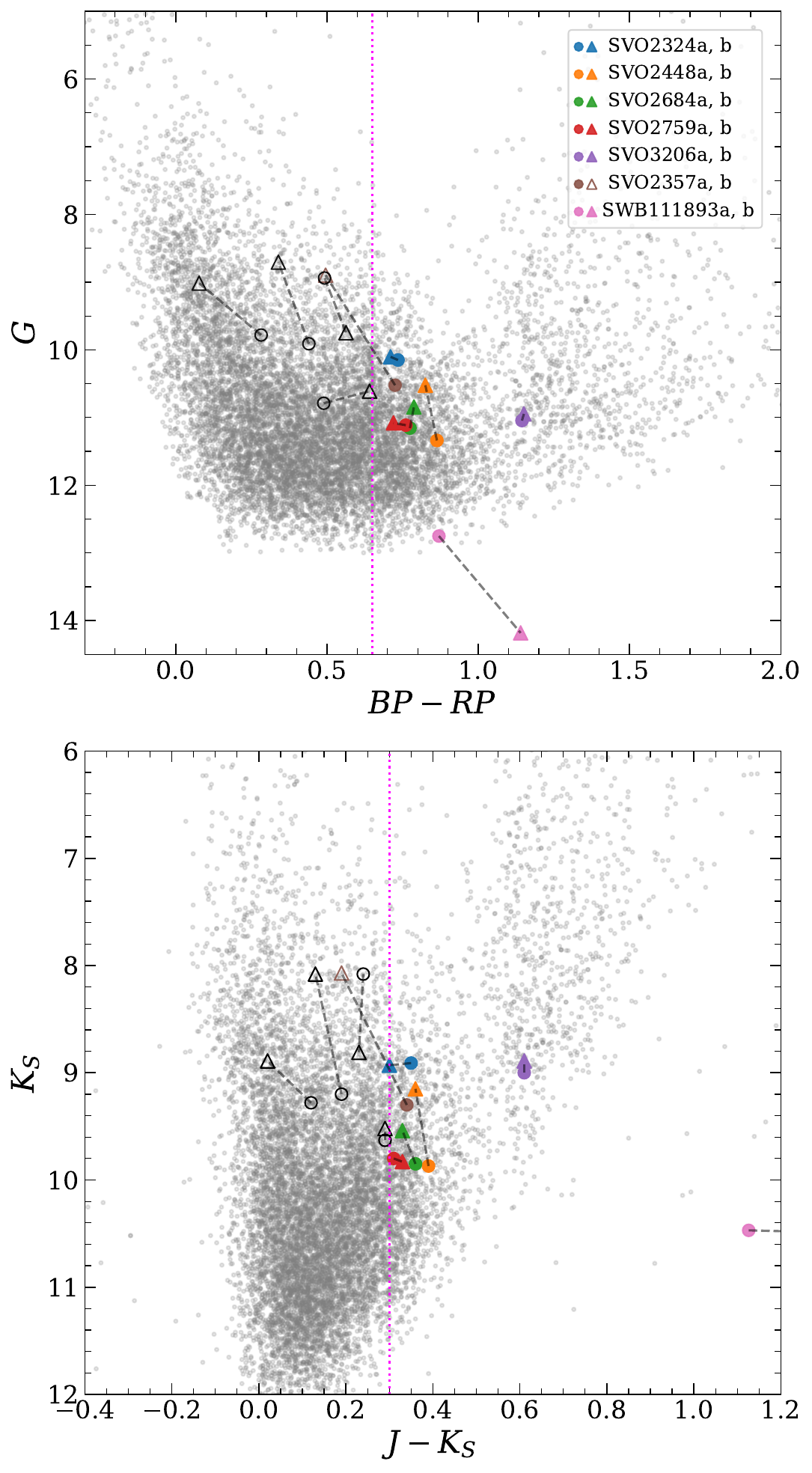} 
     \caption{
     Observed stars in the Gaia (upper panel) and 2MASS (bottom panel) CMDs, together with co-moving pair candidates of \citet[][grey dots]{Jimenez-Esteban2019}.
     The observed stars are connected with dashed lines for each pair. 
     The SWB111893 pair is apart from the other pairs because this target was selected from the catalog of \citet[][see Section~\ref{sec:sub:target}]{Hartman2020}.
     Open symbols indicate the stars for which chemical abundance measurements were not feasible.
     In the case of SVO2357 pair, spectral analysis was only available for the SVO2357a star. 
     The vertical magenta dotted line in each panel represents the approximate limit for the detailed chemical abundance measurements from the NIR spectroscopy ($BP-RP$ $=$ 0.65 and $J-K_{s}$ $=$ 0.3).
     }
     \label{fig:cmd}
\end{figure}

\begin{table*}
\scriptsize
\setlength{\tabcolsep}{4pt}
\caption{Target information from Gaia DR3.}
\label{tab:target} 
\centering                                    
\begin{tabular}{ccccccccccc}   
\hline\hline        
\multirow{2}{*}{ID} & \multirow{2}{*}{Gaia DR3 source ID}   & $\alpha$ (J2000)  & $\delta$ (J2000)  & Parallax  & $\mu_{\alpha}$    & $\mu_{\delta}$    & $G$   & RV            & Ang. Sep      & Phy. Sep \\
                    &                                   & [deg]             & [deg]             & [mas]     & [mas yr$^{-1}$]   & [mas yr$^{-1}$]   & [mag] & [km~s$^{-1}$] & [$''$]        & [AU] \\
\hline                                   
SVO2324a & 5181911066628231808 &  47.541109 &  $-$5.306764 & 5.6447 &  23.856 &   1.458 & 10.15 &  19.32 &\multirow{2}{*}{ 128.83 } & \multirow{2}{*}{   22789 } \\
SVO2324b & 5181912853334626304 &  47.542419 &  $-$5.271002 & 5.6619 &  23.679 &   1.651 & 10.10 &  19.08 & & \\
SVO2448a & 2624616237837226368 & 339.730651 &  $-$5.249625 & 4.5301 &  35.309 & $-$50.395 & 11.34 & $-$32.22 &\multirow{2}{*}{  98.86 } & \multirow{2}{*}{   21843 } \\
SVO2448b & 2623865584928390784 & 339.756399 &  $-$5.259463 & 4.5223 &  35.225 & $-$50.420 & 10.52 & $-$31.78 & & \\
SVO2684a & 6541579171444700160 & 347.030707 & $-$43.970696 & 3.4188 &  45.947 & $-$11.790 & 11.16 & $-$25.90 &\multirow{2}{*}{  63.56 } & \multirow{2}{*}{   18601 } \\
SVO2684b & 6541579201508550272 & 347.053062 & $-$43.963426 & 3.4152 &  46.094 & $-$11.827 & 10.84 & $-$25.13 & & \\
SVO2759a & 3210886042610196480 &  80.579055 &  $-$3.385248 & 3.1619 & $-$25.700 &  17.009 & 11.11 &  24.60 &\multirow{2}{*}{ 302.15 } & \multirow{2}{*}{   95124 } \\
SVO2759b & 3210885119193585536 &  80.546641 &  $-$3.462690 & 3.1909 & $-$25.726 &  16.973 & 11.08 &  24.35 & & \\
SVO3206a & 5169221839154806400 &  49.654381 &  $-$7.287289 & 2.3761 &   4.853 &  $-$8.201 & 11.04 &  32.42 &\multirow{2}{*}{  31.01 } & \multirow{2}{*}{   13001 } \\
SVO3206b & 5169221912169592832 &  49.650359 &  $-$7.279655 & 2.3939 &   4.949 &  $-$8.240 & 10.95 &  32.58 & & \\
\hline
SWB111893a & 5939115003525132800 &  254.672836 & $-$46.823396 & 1.8932 & 9.271 & $-$71.435 & 12.75 &  --          &\multirow{2}{*}{ 4.69 } & \multirow{2}{*}{   2422 } \\
SWB111893b & 5939115003525132928 &  254.671031 & $-$46.822978 & 1.9843 & 9.294 & $-$71.547 & 14.18 &  $-$191.64 & & \\
\hline                                             
\end{tabular}
\end{table*}

\subsection{IGRINS observations}\label{sec:sub:obs}
The high-resolution NIR spectroscopy for our targets was conducted using the IGRINS instrument at the Gemini-South telescope under Program GS-2021B-Q-310 and GS-2023A-Q-309 (PI: Seungsoo Hong).
These observations were carried out as part of the K-GMT science program. 
We note that although observing time was also allocated for the 2022A semester (GS-2022A-Q-219), no observations were conducted during that period. 
The IGRINS instrument consists of two separate spectrograph arms, covering the H and K bands, respectively, with a spectral resolving power of R $\sim$ 45,000 \citep{Park2014, Mace2018}.
The observations were performed in service mode over nine nights between August and November 2021 and one night in April 2023 under Band 3 condition (Image Quality $\sim$ 85, Cloud Cover $\sim$ 80). 
Each spectrum was obtained from an ABBA nod sequence observation with an estimated exposure time to achieve a signal-to-noise ratio (S/N) $\sim$ 100, as referred from the IGRINS website\footnote{https://sites.google.com/site/igrinsatgemini}. 
However, the derived S/N, estimated from the variances of each spectrum, varied from 60 to 120 depending on the star.

Out of the 20 stars observed during the 2021B semester, the spectroscopic analysis could not be performed on 9 stars. 
This was primarily due to their weak or unobservable absorption features in the NIR region, which can be attributed to the high temperature of their stellar atmospheres. 
In Figure~\ref{fig:cmd}, the observed targets are plotted on the color-magnitude diagrams (CMDs), with open symbols indicating the stars for which chemical abundance measurements were not feasible. 
It is evident from the diagram that the study of stellar chemical abundances using NIR spectroscopy (H- and K-bands) is not suitable for stars with ($BP-RP$ $>$ 0.65) or ($J-K_{s}$ $>$ 0.3).
Therefore, the five pairs, including SVO2357, for which spectroscopic analysis was not available, have been excluded from our analysis (see Figure~\ref{fig:cmd}).

The target IDs are derived from the original catalogs using specific naming conventions. 
For the targets selected from \citet{Jimenez-Esteban2019}, the IDs consist of the prefix `SVO' followed by the group ID. 
Similarly, for targets from \citet{Hartman2020}, the IDs include the prefix `SWB' followed by the catalog ID.
In Table~\ref{tab:target}, we provide the IDs, coordinates, parallax, proper motions, magnitudes, and radial velocities (RVs) for the targets, updated by the recent Gaia DR3. 
In addition, the angular and physical separations between the component stars listed in the table have been re-estimated using Gaia DR3, whereas the original catalogs were based on Gaia DR2.


\subsection{Data reduction and radial velocity measurements}\label{sec:sub:data}
The 1D spectrum was extracted using the IGRINS Pipeline Package \citep[PLP;][]{Lee2017}.
This process involved flat fielding correction, subtraction of the sky background, and wavelength calibration using OH emission and telluric lines.
To create a single continuous spectrum spanning from 1.5 to 2.4 $\mu m$, we merged the spectra from 28 and 26 echelle orders for the H and K arms, respectively, leaving a gap between 1.81 to 1.94 $\mu m$.
The final spectra were obtained after normalizing the continuum and converting the wavelengths from vacuum to air using the {\em specutils} package of {\em Astropy} \citep{AstropyCollaboration2013, AstropyCollaboration2018}.
A more detailed description of the data reduction process can be found in \citet{Lim2022}. 

Then, we measured RV of each star using cross-correlation with the synthetic spectrum from the Pollux database \citep{Palacios2010} through the IRAF {\em RV} package. 
The heliocentric RV (RV$_{helio}$), derived through the {\em rvcorrect} task, and the corresponding measurement error for each star are provided in Table~\ref{tab:param}.
It is important to note that the RVs for these stars have been updated in Gaia DR3 after our observations (see Table~\ref{tab:target}).
The RVs obtained from Gaia DR3 and our measurements demonstrate good agreement, with a typical difference of $\sim$0.3~km~s$^{-1}$, except SWB111893b. 
Furthermore, the RV discrepancies between the component stars of each target pair, as estimated by both our measurements and Gaia, are found to be smaller than 0.8~km~s$^{-1}$. 
This suggests that these paired stars share a common proper motion in 3D space, indicating a current or past physical binding in wide binary systems. 
However, a more detailed kinematic analysis is necessary to further investigate the orbital properties of these co-moving pairs.
For a simple task, we estimated the future trajectories of our samples within the Milky Way based on their current position and 3D motions using the {\em Gala} package \citep{Price-Whelan2017}.  
The component stars of SVO2448 and SVO3206 will move closer together, while the stars in the other pairs will move farther apart. 

However, in the case of SWB111893 pair, the two component stars exhibit a significant difference in our RV estimate beyond the measurement uncertainties.
Although Gaia DR3 provides RV only for the SWB111893b star ($-$191.64~km~s$^{-1}$), it is also in disagreement with our estimate ($-$58.12~km~s$^{-1}$), but it aligns with the RV of SWB111893a ($-$189.91~km~s$^{-1}$).
We suspect that the RV of SWB111893b obtained from Gaia DR3 is contaminated by the nearby SWB111893a due to their small angular separation of only $\lesssim$ 5$''$. 
The similarity in proper motion between the two stars is also doubtful due to relatively large measurement error in parallax and proper motion in the R.A. and Dec. directions, which are 0.06~mas, 0.08~mas~yr$^{-1}$, and 0.06~mas~yr$^{-1}$ for SWB111893a.  
This is in contrast to the typical errors observed in other samples, 0.02~mas in parallax and 0.02~mas~yr$^{-1}$ in proper motion for both directions. 
Even if their proper motions are accurate, it is difficult to consider them as co-moving system due to the substantial difference in space motion.
SWB111893a has ($U$, $V$, $W$) velocities of ($-$51.9, $-$226.8, $-$120.3)~km~s$^{-1}$, while SWB111893b has ($-$91.6, $-$100.8, $-$120.5)~km~s$^{-1}$, which dynamically places them in the halo and thick disk, respectively.  
A more precise understanding of their motion and orbit will be necessary once more accurate astrometric solution becomes available in future Gaia data releases.
Furthermore, in addition to this substantial kinematic discrepancy between the two stars, their spectra also display notably distinct absorption features. 
These differences indicate that SWB111893 is not a wide binary, and therefore, we have excluded them from our analysis as they are not suitable for our intended purpose. 

\begin{table}
\scriptsize
\setlength{\tabcolsep}{3pt}
\caption{Heliocentric RV and atmospheric parameters.}
\label{tab:param} 
\centering                                    
\begin{tabular}{cccccc}  
\hline\hline 
\multirow{2}{*}{ID} & RV$_{helio}$  & T$_{\rm eff}$ & $\log{g}$ & $\xi_{t}$     & [Fe/H]$_{model}$  \\ 
                    & [km~s$^{-1}$] & [K]           & [dex]     & [km~s$^{-1}$] & [dex]             \\
\hline
SVO2324a     & 19.58$\pm$0.67       & 5883 & 4.19 & 1.04 & 0.18 \\
SVO2324b     & 19.13$\pm$0.67       & 6113 & 4.26 & 1.12 & 0.18 \\ 
SVO2448a     & $-$31.33$\pm$0.24    & 5683 & 4.34 & 0.87 & 0.22 \\ 
SVO2448b     & $-$31.51$\pm$0.43    & 5804 & 4.10 & 1.03 & 0.28 \\ 
SVO2684a     & $-$25.40$\pm$0.37    & 5738 & 4.17 & 0.97 & 0.22 \\ 
SVO2684b     & $-$25.07$\pm$0.33    & 5860 & 4.05 & 1.08 & 0.29 \\ 
SVO2759a     & 24.62$\pm$1.15       & 6091 & 4.15 & 1.21 & $-$0.43 \\ 
SVO2759b     & 24.02$\pm$1.05       & 5998 & 4.13 & 1.18 & $-$0.51 \\ 
SVO3206a     & 32.56$\pm$0.29       & 4896 & 3.22 & 1.03 & $-$0.02 \\ 
SVO3206b     & 33.03$\pm$0.35       & 4896 & 3.19 & 1.04 & $-$0.02 \\ 
\hline
SWB111893a  & $-$189.91$\pm$2.24    & -- & -- & -- & -- \\ 
SWB111893b  & $-$58.12$\pm$1.03     & -- & -- & -- & -- \\ 
\hline
\end{tabular}
\end{table}

\section{Spectroscopic analysis}\label{sec:spec}
\subsection{Atmospheric parameters }\label{sec:sub:atm}
To determine the atmosphere parameters, effective temperature (${\rm T_{eff}}$), surface gravity ($\log{g}$), and microturbulence ($v_{t}$) for each star, we employed canonical photometric methods.  
It is important to mention that, in the case of NIR spectroscopy, the spectroscopic determination of atmosphere parameters through the ionization equilibrium between \ions{Fe}{i} and \ions{Fe}{ii} abundances is limited due to the absence of \ions{Fe}{ii} lines. 
In our previous study employing IGRINS data, we derived ${\rm T_{eff}}$ and $\log{g}$ for giant stars using the line-depth ratio (LDR) and equivalent width (EW) of the CO band \citep{Lim2022}. 
However, these methods are not applicable in the present study as the LDR method for the NIR region has not been validated for subdwarf stars, and the CO feature is not observable in our target stars due to their high ${\rm T_{eff}}$. 

Therefore, we determined the photometric ${\rm T_{eff}}$ of our target stars using the $(J-K_{s})$ color from the 2MASS catalog \citep{Skrutskie2006} and the color-temperature relations of \citet{GonzalezHernandez2009}. 
Since this relation includes a metallicity term, we assumed solar metallicity ([Fe/H] = 0.0 dex) as an initial value for all stars in the first iteration. 
Although we also measured ${\rm T_{eff}}$ using the relation of \citet{Mucciarelli2021} for Gaia photometric data, we adopted ${\rm T_{eff}}$ obtained from $(J-K_{s})$ in our analysis due to its lower susceptibility to extinction. 
The reddening factors for each star were obtained from various sources of \citet{Schlegel1998}, \citet{Schlafly2011}, Bayestar~17 \citep{Green2018}, Bayestar~19 \citep{Green2019}, and Stilism \citep{Lallement2018}.
Although these values were generally similar, we finally adopted E(B-V) value from \citet{Schlafly2011}, along with their extinction coefficients (R$_{J}$ = 0.723; R$_{H}$ = 0.460; R$_{K_{s}}$ = 0.310), which are available for all sample stars.
However, for the SVO2759 pair, we utilized E(B-V) values obtained from Stilism due to the significant variations among reddening sources for these stars, which could lead to substantial differences in chemical abundance measurements (for more details, see Section~\ref{sec:sub:result_Fe}).

With the initial determination of ${\rm T_{eff}}$, we calculated $\log{g}$ using the canonical relation:
\begin{eqnarray} \label{eq:logg}
\log{g_{*}} = \log{g_{\odot}} + \log{\frac{M_{*}}{M_{\odot}}} 
+ 4\log{\frac{\rm T_{eff,*}}{\rm T_{eff,\odot}}} \nonumber \\
+ 0.4(M_{bol,*} - M_{bol,\odot}), \nonumber
\end{eqnarray}
where $\log{g_{\odot}}$ = 4.44 dex, $T_{\rm eff,\odot}$ = 5777 K, and $M_{bol,\odot}$ = 4.74 for the Sun.
The bolometric correction for each star was computed using the code provided by \citet{Casagrande2018}, and stellar mass and distance were obtained from Gaia. 
We then estimated $v_{t}$ from the relation provided by \citet{Mashonkina2017}: 
\small
\begin{eqnarray} \label{eq:vt}
v_{t} = 0.14 - 0.08 \times {\rm [Fe/H]} + 4.90 \times ({\rm T_{eff}}/10^{4}) - 0.47 \times \log{g}. \nonumber 
\end{eqnarray}
\normalsize

After the first iteration of the above procedure, we obtained the second estimate of metallicity using the newly derived ${\rm T_{eff}}$, $\log{g}$, and $v_{t}$ parameters. 
We repeated the whole process with the newly derived metallicity until the output matched the input.
The final values of ${\rm T_{eff}}$, $\log{g}$, $v_{t}$, and [Fe/H] for the atmospheric models are provided in Table~\ref{tab:param}.

\begin{figure}
\centering
   \includegraphics[width=0.45\textwidth]{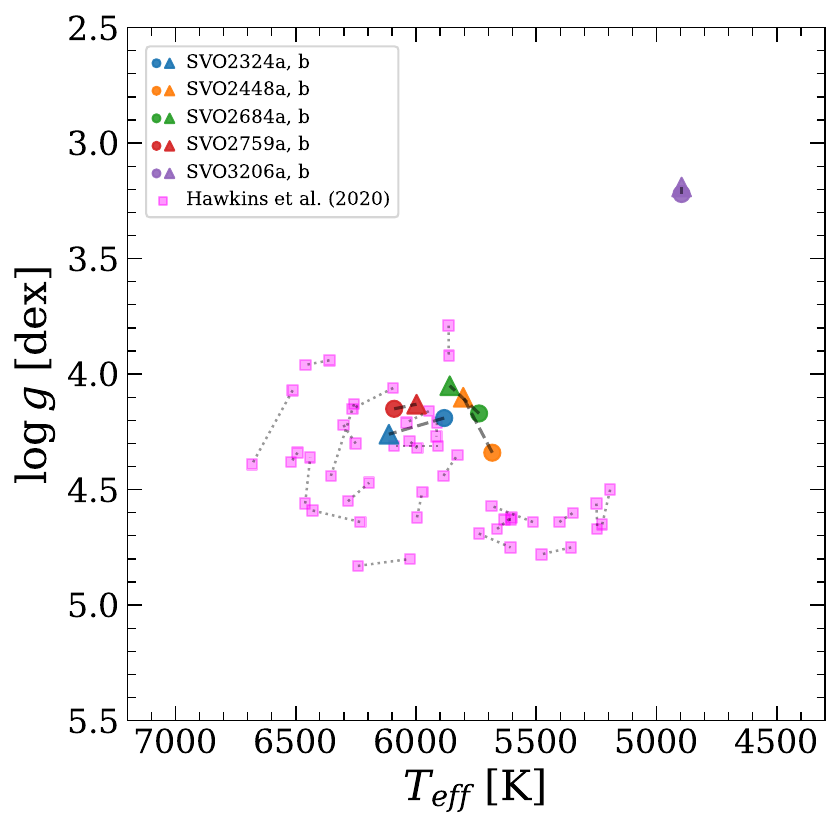} 
     \caption{
      Target stars on a Kiel diagram, together with wide binaries of \citet[][magenta squares]{Hawkins2020}.  
      Our sample pairs share similar ${\rm T_{eff}}$ and $\log{g}$ parameters between the component stars, which are comparable to the samples of \citet{Hawkins2020}. 
      In particular, two stars of the SVO3206 pair, which is the only subgiant pair among our samples, are located in almost the same region in this diagram. 
     }
     \label{fig:kiel}
\end{figure}

In Figure~\ref{fig:kiel}, we present the Kiel diagram showing the ${\rm T_{eff}}$ and $\log{g}$ values for our target stars, along with the connection between each pair of stars.
As shown in this figure, SVO3026a,~b are identified as subgiant stars, while the others are subdwarf stars.  
It is important to note that while the component stars of our wide binary samples happen to be located at the same evolution stage, wide binaries comprising different types of stars, such as main-sequence and white dwarf pairs, are also reported in the literature \citep[e.g.,][]{El-Badry2021, Zhao2023}.
Furthermore, the component stars of each wide binary demonstrate similarities in all atmospheric parameters, with differences of $\lesssim$ 200~K in ${\rm T_{eff}}$, $\lesssim$ 0.2~dex in $\log{g}$, and $\lesssim$ 0.1~km~s$^{-1}$ in $v_{t}$. 
In particular, SVO3206a and SVO3206b exhibit almost identical characteristics, with $\Delta {\rm T_{eff}}$ = 0~K, $\Delta \log{g}$ = 0.03~dex, and $\Delta v_{t}$ = 0.01~km~s$^{-1}$. 
These similarities suggest that the two stars in each wide binary were formed together and have evolved equally so far. 

On the other hand, since SVO3206a and b are subgiant stars, we were able to measure ${\rm T_{eff}}$ using the LDR relations suggested by \citet{Fukue2015}, as utilized in \citet{Lim2022}. 
The LDR ${\rm T_{eff}}$ is estimated to be 4980~K for both stars, which is only 84~K larger than the value derived from the $(J-K_{s})$ color. 
In addition, ${\rm T_{eff}}$ values measured using the relation with the EW of the CO-overtone band at 2.293~$\mu m$, suggested by \citet{Park2018}, are also close to the other estimations, as 4898~K for SVO3206a and 4891~K for SVO3206b. 
The consistency of ${\rm T_{eff}}$ values obtained from three different methods further supports the reliability of our parameter determination. 

\subsection{Chemical abundance analysis}\label{sec:sub:abund}
The chemical element abundances of each star were determined using the spectral synthesis method with the {\em synth} driver of the 2019NOV version of the local thermodynamic equilibrium (LTE) code MOOG \citep{Sneden1973}.  
Model atmospheres were constructed by interpolating a grid of plane-parallel MARCS models \citep{Gustafsson2008} based on the parameters determined in the previous section. 
The chemical abundances of Fe, Na, Mg, Al, Si, S, K, Ca, Ti, Cr, Ni, and Ce were measured from absorption lines listed in \citet{Lim2022}, referring \citet{Afsar2018}. 
In addition, atomic and molecular lines generated using the recent version of {\em linemake}\footnote{https://github.com/vmplacco/linemake} \citep{Placco2021} were used to make synthetic spectra. 
The chemical abundances were measured from each absorption line, and the final values were determined as the mean of the measurements for each element. 

For the SVO3206a and SVO3206b stars, we were able to measure the abundances of C, N, and O from CO, CN, and OH features, respectively.
However, these molecular features were not observed in other stars due to their high ${\rm T_{eff}}$. 
The abundances of O, C, and N were sequentially measured, and the measurements were repeated with the updated abundances in the model atmosphere since these elements are tied in molecules.
The derived abundances for SVO3206a and SVO3206b are as follows: [C/Fe] = $-$0.139; $-$0.168 dex, [N/Fe] = +0.030; +0.121 dex, and [O/Fe] = $-$0.021; +0.011 dex, respectively. 

\begin{table*}[ht]
\scriptsize
\setlength{\tabcolsep}{1.8pt}
\caption{Chemical abundance ratios for Fe, Na, Mg, Al, and Si elements.}
\label{tab:abund} 
\centering
\begin{tabular}{@{\extracolsep{2pt}}c r r r  r r r  r r r  r r r  r r r  r r r  r r r  r r r }   
\hline\hline 
\multirow{2}{*}{ID}	& \multicolumn{3}{c}{Fe}	& \multicolumn{3}{c}{Fe$_{\rm NLTE}$}	& \multicolumn{3}{c}{Na}	& \multicolumn{3}{c}{Mg}
				& \multicolumn{3}{c}{Mg$_{\rm NLTE}$}	& \multicolumn{3}{c}{Al}	& \multicolumn{3}{c}{Si}	& \multicolumn{3}{c}{Si$_{\rm NLTE}$} \\
\cline{2-4} \cline{5-7} \cline{8-10} \cline{11-13} \cline{14-16} \cline{17-19} \cline{20-22} \cline{23-25}
                                  & [X/H] & $\sigma$ & $N$ & [X/H] & $\sigma$ & $N$ & [X/H] & $\sigma$ & $N$ & [X/H] & $\sigma$ & $N$ 
                                  & [X/H] & $\sigma$ & $N$ & [X/H] & $\sigma$ & $N$ & [X/H] & $\sigma$ & $N$ & [X/H] & $\sigma$ & $N$ \\
\hline   
SVO2324a &  0.18 &  0.05 & 17 &  0.18 &  0.04 & 12 &  0.17 &  0.05 &  4 & $-$0.03 &  0.08 &  9 & $-$0.03 &  0.08 &  9 &  0.25 &  0.15 &  5 &  0.20 &  0.08 & 12 &  0.19 &  0.08 & 12 \\
SVO2324b &  0.18 &  0.07 & 16 &  0.18 &  0.07 & 12 &  0.15 &  0.08 &  4 &  0.06 &  0.07 & 10 &  0.06 &  0.07 & 10 &  0.19 &  0.14 &  5 &  0.17 &  0.08 & 11 &  0.15 &  0.08 & 11 \\
SVO2448a &  0.22 &  0.08 & 23 &  0.22 &  0.08 & 17 &  0.37 &  0.01 &  3 &  0.02 &  0.07 & 10 &  0.02 &  0.07 & 10 &  0.35 &  0.17 &  5 &  0.27 &  0.07 & 12 &  0.26 &  0.07 & 12 \\
SVO2448b &  0.28 &  0.06 & 20 &  0.28 &  0.06 & 15 &  0.36 &  0.03 &  3 &  0.15 &  0.06 & 10 &  0.15 &  0.07 & 10 &  0.40 &  0.17 &  5 &  0.33 &  0.06 & 11 &  0.32 &  0.06 & 11 \\
SVO2684a &  0.22 &  0.06 & 22 &  0.23 &  0.06 & 17 &  0.35 &  0.07 &  3 &  0.05 &  0.08 & 10 &  0.05 &  0.09 & 10 &  0.31 &  0.20 &  5 &  0.28 &  0.11 & 11 &  0.26 &  0.11 & 11 \\
SVO2684b &  0.29 &  0.10 & 26 &  0.29 &  0.10 & 20 &  0.46 &  0.09 &  3 &  0.15 &  0.06 & 10 &  0.15 &  0.07 & 10 &  0.40 &  0.18 &  5 &  0.32 &  0.07 & 11 &  0.30 &  0.07 & 11 \\
SVO2759a & $-$0.43 &  0.08 & 17 & $-$0.43 &  0.07 & 12 & $-$0.49 &  0.06 &  4 & $-$0.46 &  0.06 &  8 & $-$0.46 &  0.06 &  8 & $-$0.20 &  0.07 &  3 & $-$0.35 &  0.05 & 12 & $-$0.36 &  0.05 & 12 \\
SVO2759b & $-$0.51 &  0.07 & 18 & $-$0.49 &  0.08 & 13 & $-$0.54 &  0.09 &  4 & $-$0.65 &  0.10 &  8 & $-$0.65 &  0.12 &  8 & $-$0.33 &  0.06 &  4 & $-$0.40 &  0.08 & 12 & $-$0.41 &  0.09 & 12 \\
SVO3206a & $-$0.02 &  0.09 & 23 & $-$0.02 &  0.08 & 17 &  0.13 &  0.03 &  3 & $-$0.00 &  0.07 & 10 & $-$0.01 &  0.08 & 10 &  0.20 &  0.24 &  5 &  0.02 &  0.06 & 12 &  0.00 &  0.06 & 12 \\
SVO3206b & $-$0.02 &  0.08 & 22 & $-$0.03 &  0.07 & 16 &  0.15 &  0.01 &  3 & $-$0.03 &  0.05 &  9 & $-$0.04 &  0.06 &  9 &  0.23 &  0.20 &  6 &  0.04 &  0.07 & 12 &  0.03 &  0.07 & 12 \\
\hline
\end{tabular}
\end{table*}
\begin{table*}
\scriptsize
\setlength{\tabcolsep}{1.8pt}
\caption{Chemical abundance ratios for S, Ca, Ti, Cr, and Ni elements.}
\label{tab:abund2} 
\centering
\begin{tabular}{@{\extracolsep{2pt}}c r r r  r r r  r r r  r r r  r r r  r r r  r r r  r r r  }   
\hline\hline 
\multirow{2}{*}{ID}	& \multicolumn{3}{c}{S}	& \multicolumn{3}{c}{Ca}	& \multicolumn{3}{c}{Ca$_{\rm NLTE}$}	& \multicolumn{3}{c}{Ti}
                                  & \multicolumn{3}{c}{Ti$_{\rm NLTE}$}	& \multicolumn{3}{c}{Cr}	& \multicolumn{3}{c}{Cr$_{\rm NLTE}$}	& \multicolumn{3}{c}{Ni} \\
\cline{2-4} \cline{5-7} \cline{8-10} \cline{11-13} \cline{14-16} \cline{17-19} \cline{20-22} \cline{23-25}
                                  & [X/H] & $\sigma$ & $N$        & [X/H] & $\sigma$ & $N$ & [X/H] & $\sigma$ & $N$ & [X/H] & $\sigma$ & $N$ 
                                  & [X/H] & $\sigma$ & $N$ & [X/H] & $\sigma$ & $N$ & [X/H] & $\sigma$ & $N$ & [X/H] & $\sigma$ & $N$ \\
\hline      
SVO2324a &  0.28 &  0.06 &  9 &  0.23 &  0.07 & 10 &  0.29 &  0.07 &  9 &  0.18 &  0.17 &  7 &  0.22 &  0.18 &  4 &  0.09 &  0.07 &  3 &  0.16 &  0.07 &  3 &  0.12 &  0.07 &  5 \\
SVO2324b &  0.20 &  0.10 &  7 &  0.20 &  0.09 &  9 &  0.25 &  0.08 &  7 &  0.12 &  0.11 &  3 &  0.21 &  0.11 &  3 &  0.12 &  0.05 &  2 &  0.20 &  0.05 &  2 &  0.20 &  0.04 &  4 \\
SVO2448a &  0.28 &  0.02 &  8 &  0.20 &  0.04 & 10 &  0.23 &  0.06 &  9 &  0.12 &  0.08 &  7 &  0.17 &  0.08 &  4 &  0.09 &  0.10 &  3 &  0.14 &  0.10 &  3 &  0.28 &  0.07 &  6 \\
SVO2448b &  0.28 &  0.05 &  7 &  0.32 &  0.08 &  9 &  0.36 &  0.06 &  8 &  0.28 &  0.12 &  7 &  0.23 &  0.07 &  3 &  0.23 &  0.14 &  2 &  0.30 &  0.14 &  2 &  0.35 &  0.08 &  5 \\
SVO2684a &  0.40 &  0.05 &  9 &  0.19 &  0.04 &  9 &  0.24 &  0.04 &  8 &  0.10 &  0.04 &  5 &  0.20 &  0.05 &  4 &  0.09 &  0.10 &  3 &  0.16 &  0.10 &  3 &  0.28 &  0.04 &  4 \\
SVO2684b &  0.32 &  0.05 &  9 &  0.28 &  0.04 &  9 &  0.35 &  0.04 &  8 &  0.20 &  0.13 &  8 &  0.27 &  0.15 &  5 &  0.23 &  0.04 &  2 &  0.30 &  0.04 &  2 &  0.33 &  0.07 &  6 \\
SVO2759a & $-$0.18 &  0.07 &  8 & $-$0.33 &  0.06 & 10 & $-$0.26 &  0.07 &  9 &   -- &   -- &  -- &   -- &   -- &  -- & $-$0.32 &  0.01 &  2 & $-$0.14 &  0.01 &  2 & $-$0.41 &  0.03 &  3 \\
SVO2759b & $-$0.13 &  0.12 &  9 & $-$0.46 &  0.07 &  8 & $-$0.40 &  0.07 &  8 &   -- &   -- &  -- &   -- &   -- &  -- & $-$0.39 &  0.00 &  1 & $-$0.21 &  0.00 &  1 & $-$0.52 &  0.09 &  3 \\
SVO3206a & $-$0.06 &  0.12 &  8 &  0.08 &  0.11 & 10 &  0.08 &  0.09 &  9 & $-$0.08 &  0.06 &  9 &  0.02 &  0.04 &  5 & $-$0.22 &  0.12 &  3 & $-$0.15 &  0.12 &  3 & $-$0.07 &  0.06 &  6 \\
SVO3206b & $-$0.07 &  0.11 &  8 &  0.09 &  0.09 & 10 &  0.09 &  0.07 &  9 & $-$0.02 &  0.07 &  9 &  0.07 &  0.06 &  5 & $-$0.15 &  0.12 &  3 & $-$0.08 &  0.12 &  3 & $-$0.02 &  0.06 &  6 \\                     
\hline
\end{tabular}
\end{table*}

Furthermore, we applied line-by-line non-LTE (NLTE) abundance corrections for Fe, Mg, Si, Ca, Ti, and Cr using previous studies by \citet{Mashonkina2007}, \citet{Bergemann2010}, and \citet{Bergemann2012, Bergemann2013, Bergemann2015} for 1D plane-parallel atmosphere models\footnote{https://nlte.mpia.de}. 
The comparison between LTE and NLTE abundances shows that the differences are very small for Fe and Mg. 
However, the NLTE effects generally lead to an increase in the abundance of Ca, Ti, and Cr elements by $\sim$0.06~dex on average, whereas the abundance of Si is decreased by $\sim$0.01~dex. 

The LTE and NLTE chemical abundances, represented as [X/H] adopting the solar scale of \citet{Asplund2009}, the line-to-line scatter ($\sigma$), and the number of lines ($N$) for each element are presented in Tables~\ref{tab:abund} and \ref{tab:abund2}. 
We note that the Ti abundance for SVO2759a,~b stars were not measured due to the absence of absorption lines caused by their high ${\rm T_{eff}}$.
The statistical measurement error for the abundance ratios can be estimated as $\sigma$/$\sqrt{N}$ for each element. 
The typical errors on the [Fe/H] measurements are less than 0.02~dex, and those on the [Mg/H] measurements are $\lesssim$ 0.03~dex. 

In addition, systematic errors in the chemical abundance measurements can arise from uncertainties in the determination of atmosphere parameters. 
To assess the impact of these systematic errors, we used Monte Carlo sampling to estimate the uncertainties in ${\rm T_{eff}}$, $\log{g}$, and $v_{t}$. 
The uncertainty in [Fe/H] was initially obtained from the statistical measurement error. 
Then, the uncertainties in ${\rm T_{eff}}$, $\log{g}$, and $v_{t}$ were computed by taking into account the errors in magnitudes, mass, and distance from Gaia and 2MASS data, along with the error in [Fe/H] and fitting error of each equation. 
The typical uncertainties for our samples were determined to be $\pm$190~K in ${\rm T_{eff}}$, $\pm$0.06~dex in $\log{g}$, $\pm$0.07~km~s$^{-1}$ in $v_{t}$, and $\pm$0.02~dex in [Fe/H].  
We then re-estimated the chemical abundances by adopting eight-atmosphere models with different parameters that varied by the uncertainties for SVO2448b, which has a median ${\rm T_{eff}}$ among the samples. 
The systematic errors for each element, presented in Table~\ref{tab:error}, were obtained by comparing the newly estimated abundances with the original values.
The abundance variations were found to be less than 0.2~dex for all elements, with the effect of ${\rm T_{eff}}$ being the most significant, as expected given the large uncertainty in ${\rm T_{eff}}$ measurement. 

\begin{table}
\caption{Systematic error due to the uncertainty in the atmosphere parameters for SVO2448b.}
\label{tab:error} 
\scriptsize
\setlength{\tabcolsep}{1.5pt}
\centering                                    
\begin{tabular}{@{\extracolsep{3pt}}c c c c c c c c c} 
\hline\hline 
\multirow{2}{*}{Species}    & \multicolumn{2}{c}{$T_{\rm eff}$} & \multicolumn{2}{c}{$\log{g}$} & \multicolumn{2}{c}{$v_{t}$}   & \multicolumn{2}{c}{[Fe/H]} \\
        & \multicolumn{2}{c}{(5804~K)}  & \multicolumn{2}{c}{(4.10~dex)}    & \multicolumn{2}{c}{(1.03~km~s$^{-1}$)}    & \multicolumn{2}{c}{($+$0.27~dex)} \\ 
        \cline{2-3} \cline{4-5} \cline{6-7} \cline{8-9} 
        & $-$208 & $+$208               & $-$0.07 & $+$0.07                 & $-$0.07 & $+$0.07                         & $-$0.01 & $+$0.01 \\
\hline
Fe      & $-$0.13 & $+$0.13 & $+$0.01 & $-$0.01 & $+$0.01 & $-$0.01 & $+$0.00 & $-$0.00 \\
Na      & $-$0.17 & $+$0.16 & $+$0.01 & $-$0.03 & $+$0.00 & $-$0.02 & $-$0.01 & $-$0.01 \\
Mg      & $-$0.16 & $+$0.16 & $+$0.02 & $-$0.02 & $+$0.01 & $-$0.01 & $+$0.00 & $+$0.00 \\
Al      & $-$0.13 & $+$0.13 & $+$0.01 & $-$0.01 & $+$0.01 & $-$0.00 & $+$0.00 & $+$0.00 \\
Si      & $-$0.10 & $+$0.11 & $+$0.01 & $-$0.02 & $+$0.01 & $-$0.01 & $-$0.00 & $-$0.00 \\
S       & $+$0.08 & $-$0.08 & $-$0.02 & $+$0.01 & $-$0.00 & $-$0.01 & $-$0.01 & $-$0.00 \\
Ca      & $-$0.15 & $+$0.14 & $+$0.01 & $-$0.01 & $+$0.01 & $-$0.01 & $-$0.00 & $-$0.00 \\
Ti      & $-$0.20 & $+$0.17 & $-$0.00 & $-$0.00 & $-$0.00 & $-$0.01 & $-$0.00 & $-$0.00 \\
Cr      & $-$0.17 & $+$0.15 & $+$0.02 & $+$0.00 & $+$0.02 & $+$0.00 & $+$0.01 & $+$0.00 \\
Ni      & $-$0.10 & $+$0.11 & $+$0.00 & $+$0.00 & $+$0.01 & $-$0.01 & $+$0.00 & $+$0.00 \\
\hline                                             
\end{tabular}
\end{table}

\subsection{Comparison with spectroscopic survey data}\label{sec:sub:comp}
The RV and chemical abundances of some sample stars were compared with data from spectroscopic surveys. 
Specifically, the data for SVO2759b were obtained from the Large Sky Area Multi-Object fiber Spectroscopic Telescope (LAMOST) medium resolution spectroscopic survey \citep[R $\sim$ 7,500;][]{Liu2020}, and the data for SVO2448a and SVO2448b were obtained from GALactic Archaeology with HERMES \citep[GALAH, R $\sim$ 28,000;][]{Buder2021}.

The estimates in this study show good agreement with the survey data.  
The differences in RV between our estimations and the survey data are less than 1~km~s$^{-1}$.
In this study, the RVs are estimated to be 24.02, $-$31.33, and $-$32.12~km~s$^{-1}$ for SVO2759b, SVO2448a, and SVO2448b, respectively, while 23.59, $-$31.51, and $-$31.66~km~s$^{-1}$ are obtained from the survey data. 

In the case of [Fe/H] ratio, our estimations and values obtained from the GALAH survey are similar, as 0.22/0.28~dex versus 0.23/0.30~dex for SVO2448a/SVO2448b.
The differences in chemical abundances for other elements between this study and GALAH for these two stars are also less than 0.1~dex, with an average difference of 0.05~dex, except for Mg. 
However, our estimations of Mg abundances are approximately 0.2~dex lower than those from GALAH in both stars. 
For SVO2759b, although the [Fe/H] value obtained from LAMOST ($-$0.35~dex) is somewhat larger than that from our study ($-$0.51~dex), this discrepancy could be attributed to the lower resolution of the LAMOST survey or uncertain atmosphere parameters determination.

\section{Chemical homogeneity of wide binary system}\label{sec:result}

\subsection{Fe abundance}\label{sec:sub:result_Fe}
As mentioned in Section~\ref{sec:intro}, previous studies have reported similar metallicity between component stars in many wide binary pairs based on optical spectroscopic observations \citep[e.g.,][]{Andrews2018, Hawkins2020}. 
In this study, we also find that the five wide binary pairs show nearly identical [Fe/H] abundance ratios between the component stars, which is consistent with the relatively similar atmospheric parameters within each pair (see Table~\ref{tab:param}). 
The differences in [Fe/H] between the component stars are 0.002, 0.059, 0.072, 0.073, and 0.001 dex for the SVO2324, SVO2448, SVO2684, SVO2759, and SVO3206 pairs, respectively.
These differences are comparable to the statistical error on the Fe abundance measurement ($\sim$0.02~dex). 
It is noteworthy that the SVO3206a and SVO3206b stars exhibit a remarkable similarity in Fe abundances, as well as equivalent ${\rm T_{eff}}$ and other similar parameters, as shown in Figures~\ref{fig:cmd} and \ref{fig:kiel}.
When considering the NLTE [Fe/H] ratios, the differences remain small with $\Delta$[Fe/H] of 0.001, 0.058, 0.062, 0.059, and 0.008 dex for each wide binary pair, respectively. 

\begin{figure}
\centering
   \includegraphics[width=0.47\textwidth]{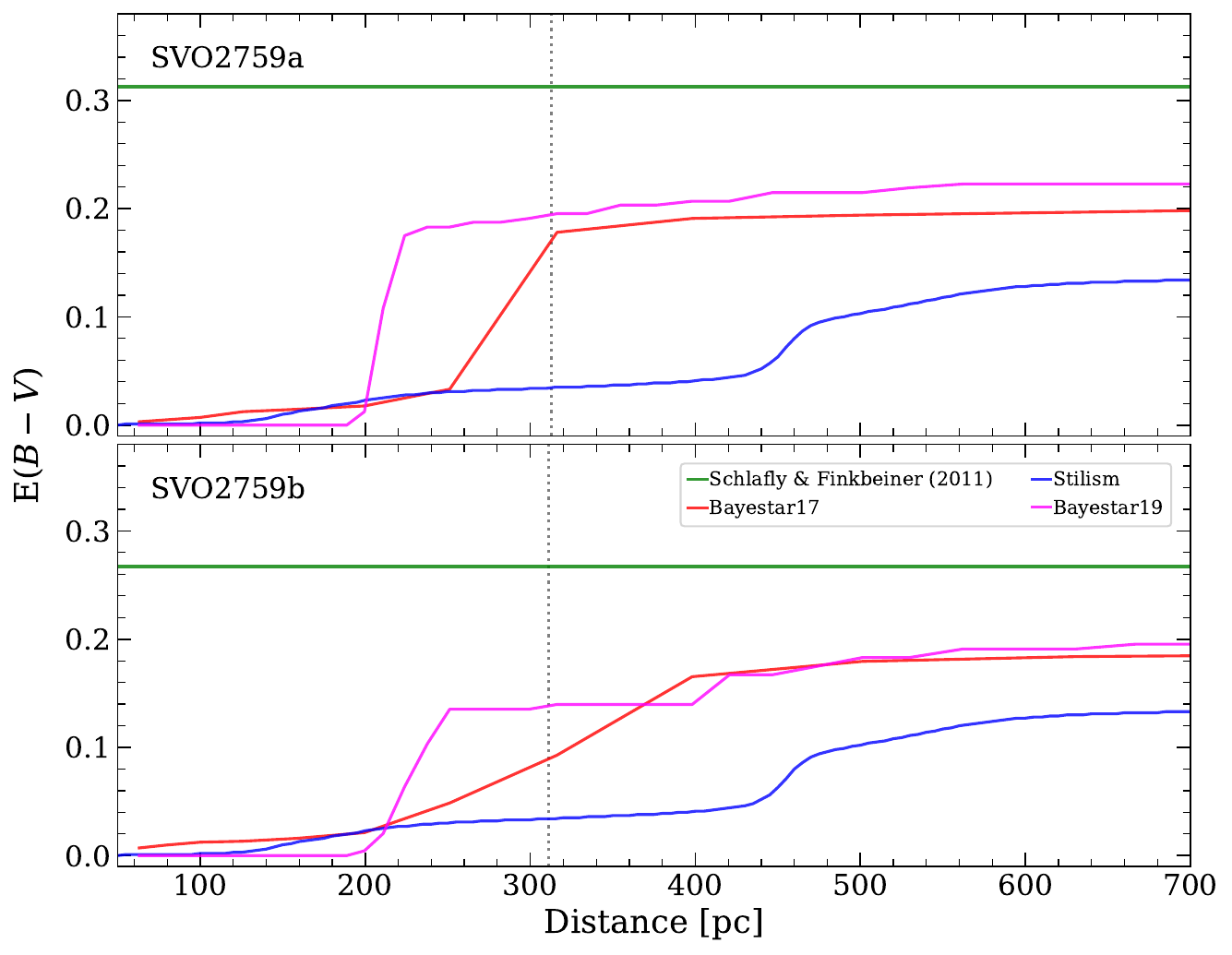} 
     \caption{
     E(B-V) values for SVO2759 pair stars obtained from \citet{Schlafly2011}, Bayestar~17 \citep{Green2018}, Bayestar~19 \citep{Green2019}, and Stilism \citep{Lallement2018}.
     Vertical dotted lines indicate the location of each star, estimated from Gaia DR3 parallaxes. 
     Reddening values from different sources are significantly but differently increased in front of or behind the target stars, while \citet{Schlafly2011} do not provide 3D reddening value.  
     }
     \label{fig:ebv}
\end{figure}

\begin{figure*}
\centering
   \includegraphics[width=0.95\textwidth]{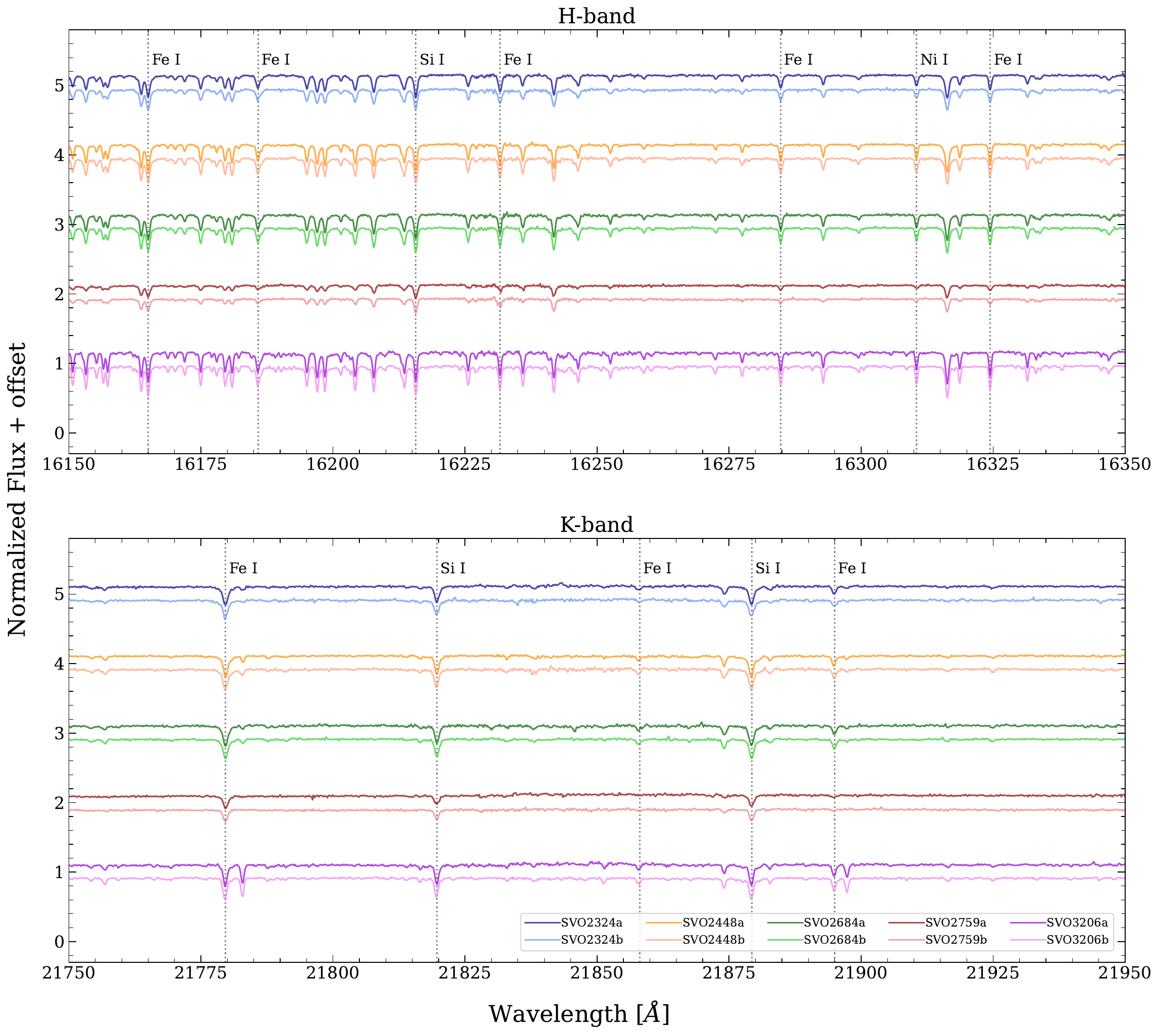} 
     \caption{
      The comparison of continuum-normalized spectra for each wide binary pair in the H- and K-band regions, including several Fe, Si, and Ni lines. 
      Two spectra of each pair demonstrate almost identical spectral features and the strength of absorption lines. 
     }
     \label{fig:spectra}
\end{figure*}

However, in the case of the SVO2759 pair, the difference in [Fe/H] between the two component stars varies with the applied reddening values. 
As discussed in Section~\ref{sec:sub:atm}, noticeable variations in E(B-V) are derived from different sources for the SVO2759a and SVO2759b stars, whereas other stars show almost similar values.
In particular, 3D extinction maps from Bayestar and Stilism demonstrate that the E(B-V) values for these two stars are highly dependent on their respective location (see Figure~\ref{fig:ebv}).
It appears that these variations in E(B-V) values may be attributed to the the two stars being located either in front of or behind a highly obscured region.
These differences in E(B-V) can lead to variations in ${\rm T_{eff}}$, which, in turn, affect the [Fe/H] measurements. 
The [Fe/H] ratios were measured as $-$0.08/$-$0.20~dex using \citet{Schlafly2011}, $-$0.20/$-$0.49~dex using Bayestar~17 \citep{Green2018}, and $-$0.43/$-$0.51~dex using Stilism \citep{Lallement2018} for the SVO2759a/SVO2759b stars, respectively.
Among these estimates, we have adopted the E(B-V) values of Stilism for the SVO2759 pair.
This choice was based on these values being the smallest, with the least difference in E(B-V) and [Fe/H] between the two stars, assuming that both stars evolved in a similar environment.
We note that the reddening values of \citet{Schlafly2011} were applied for other stars, because the Stilism reddening map is not available for distant stars.

Figure~\ref{fig:spectra} displays a comparison of the observed spectra of component stars in each wide binary pair on the H- and K-band regions. 
As shown in this figure, the two stars comprising a wide binary system exhibit remarkably similar features throughout the entire spectral region, including the absorption lines of Fe, Si, and Ni.   
The fact that the observed spectra of the component stars overlap so closely reinforces that our target stars are not randomly co-moving pairs but rather widely separated binary systems. 
Furthermore, the striking similarity in the spectral features of the two stars with similar atmospheric and chemical properties 
demonstrates the stability and effectiveness of the NIR spectroscopic observation using IGRINS and the data reduction technique utilized in this study. 

\begin{figure}
\centering
   \includegraphics[width=0.47\textwidth]{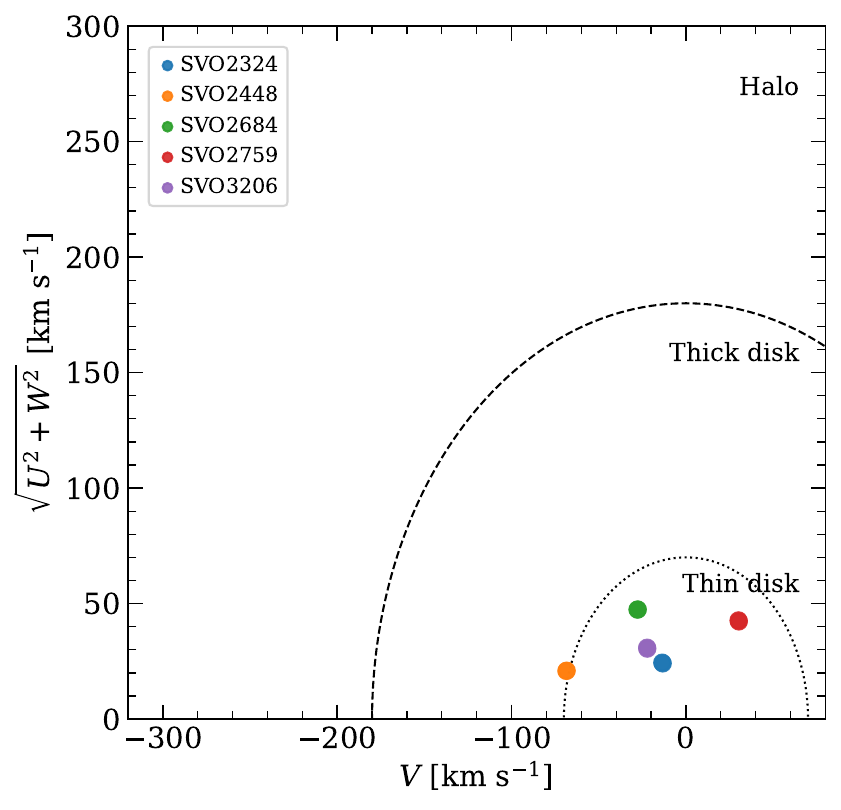} 
     \caption{
     Toomre diagram for our wide binary samples.
     The curved dashed and dotted lines represent total velocities of 180~km~s$^{-1}$ and 70~km~s$^{-1}$, which divide the halo, thick disk, and thin disk.
     Our samples dynamically belong to the thin disk. 
     }
     \label{fig:toomre}
\end{figure}

The metallicity of our samples ranges from $-$0.5~dex to $+$0.3~dex in terms of [Fe/H] ratio, which falls within the range where a large number of wide binaries are reported in the literature \citep[see][]{Hwang2021}.
Most of the known wide binaries within this metallicity range are associated with the Galactic disk \citep[e.g.,][]{Andrews2018, Hawkins2020}.
Similarly, our five wide binary samples are also dynamically associated with the thin disk, as shown in Figure~\ref{fig:toomre}, with a vertical distance ($Z_{max}$) of less than 0.5~kpc.
Therefore, these samples can be considered typical metal-rich disk wide binaries. 
While \citet{Hwang2021} suggested that the plenty of wide binaries in this metallicity range may be influenced by radial migration of stars, drawing a firm conclusion about their formation mechanism is challenging.

\subsection{Other elements}\label{sec:sub:result_other}
\begin{figure}[t]
\centering
   \includegraphics[width=0.48\textwidth]{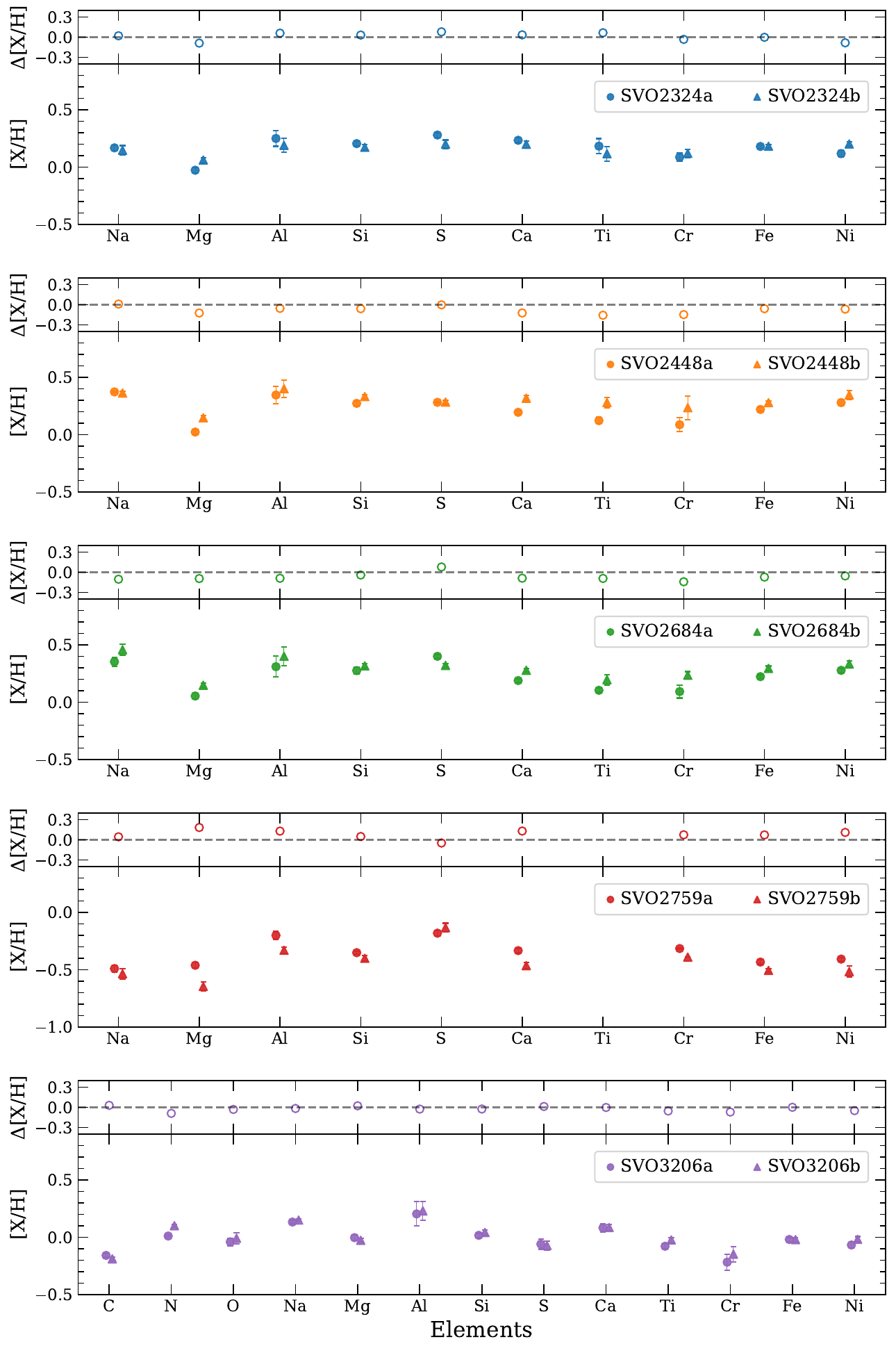} 
     \caption{
     [X/H] abundance ratios and their measurement error ($\pm$1$\sigma$) for each wide binary system and their differences between component stars ($\Delta$[X/H]).
     The chemical abundances of 10 elements are plotted for four wide binary pairs, while 13 elements, including C, N, and O, are indicated for the SVO3206 pair. 
     The dotted lines in the $\Delta$[X/H] plot indicate the value of 0.0~dex.
     }
     \label{fig:abund}
\end{figure}

Figure~\ref{fig:abund} presents the abundance ratios and measurement errors for each element in terms of [X/H], together with the differences between the component stars of each wide binary system. 
The typical measurement error is less than 0.03~dex for all elements, except for Al and Cr, where it is somewhat larger at 0.07 and 0.05~dex, respectively.
Our sample stars do not show any peculiar chemical abundance patterns compared to the general field stars in the Milky Way.
However, our measurements of the [Mg/H] abundance ratio appear to be underestimated when compared to the other elements, as suspected from the comparison with GALAH data (see Section~\ref{sec:sub:comp}).
Further investigation for the origin of this underestimation from a larger sample is necessary, while the systematic uncertainties by the absorption line information or our measurements are speculated. 

Regarding the abundance differences within each wide binary system, as anticipated from the small differences in [Fe/H] and the spectral similarities shown in Figure~\ref{fig:spectra}, the two component stars exhibit almost identical chemical compositions for all the elements measured in this study.
The average abundance difference among all elements is ranged from 0.05 to 0.09~dex for SVO2324, SVO2448, SVO2684, and SVO2759 pairs, while that for the SVO3206 pair is 0.03~dex. 
In the case of the SVO3206 pair, [C, N, O/H] abundance ratios are also indicated in the lower panel of Figure~\ref{fig:abund}, with differences between SVO3206a and SVO3206b estimated to be 0.029, 0.091, and 0.032 dex for C, N, and O elements, respectively.
These small differences are consistent with those for the other elements within the same pair.

The small differences in all abundance ratios, as well as the comparable atmospheric parameters, between the two stars in each wide binary indicate that they formed simultaneously in the same environment and undergo similar evolutionary processes.
Thus, although the two stars are currently separated by a large distance ($>$ 10,000~AU), they were likely formed simultaneously in close proximity or from a single, large, homogeneous gas cloud. 
This aligns with various scenarios proposed for the formation of wide binary systems, as introduced in Section~\ref{sec:intro}.

Furthermore, the $\Delta$[X/H] values for the sample wide binaries are comparable to those estimated in other studies based on high-resolution optical spectroscopy with R $\gtrsim$ 45,000 \citep[$\Delta \rm{[X/H]}$ = 0.05 $\sim$ 0.10 dex;][]{Hawkins2020, Nelson2021}.
This suggests that high-resolution NIR spectroscopy can effectively conduct a detailed chemical abundance study with a comparable level of reliability to typical optical spectroscopy. 
In addition, assuming the identical chemical properties of wide binary components, our results support the notion that IGRINS can be used for the chemical tagging of stars in the Milky Way with an accuracy of $\lesssim$0.1~dex.

\section{Trend of abundance differences}\label{sec:trend}
\subsection{Trend with binary separation}\label{sec:sub:trend_1}
Most wide binary pairs, including the ones in this study, show a homogeneous chemical composition between the component stars. 
However, determining the origin of wide binary pairs solely based on this chemical similarity is challenging, as multiple scenarios have been suggested based on this characteristic \citep[e.g.,][]{Kouwenhoven2010, Elliott2016}. 
The possibility of various formation channels depending on the wide binary is also one of the main tasks in understanding their origin. 

To better understand the origin of wide binary, \citet{Hwang2021} suggested examining the mass ratios and orbital eccentricity of wide binary systems as a function of metallicity for a large sample. 
In addition, \citet{Ramirez2019} proposed that studying the correlation between the differences in chemical abundances and the separation between the component stars would provide important insights into the wide binary formation scenarios. 
They demonstrated that the absolute differences in chemical abundances tend to increase with increasing separation, and this trend has also been reported by \citet{Liu2021}.
This observed trend can be attributed to the notion that larger star-forming clouds may be less homogeneous compared to smaller clouds, assuming that stars in a wide binary pair are formed from a single cloud. 
The larger differences in chemical abundances in wide binaries with larger separations may arise from the local inhomogeneities within the cloud during the star formation process.

\begin{figure}
\centering
   \includegraphics[width=0.48\textwidth]{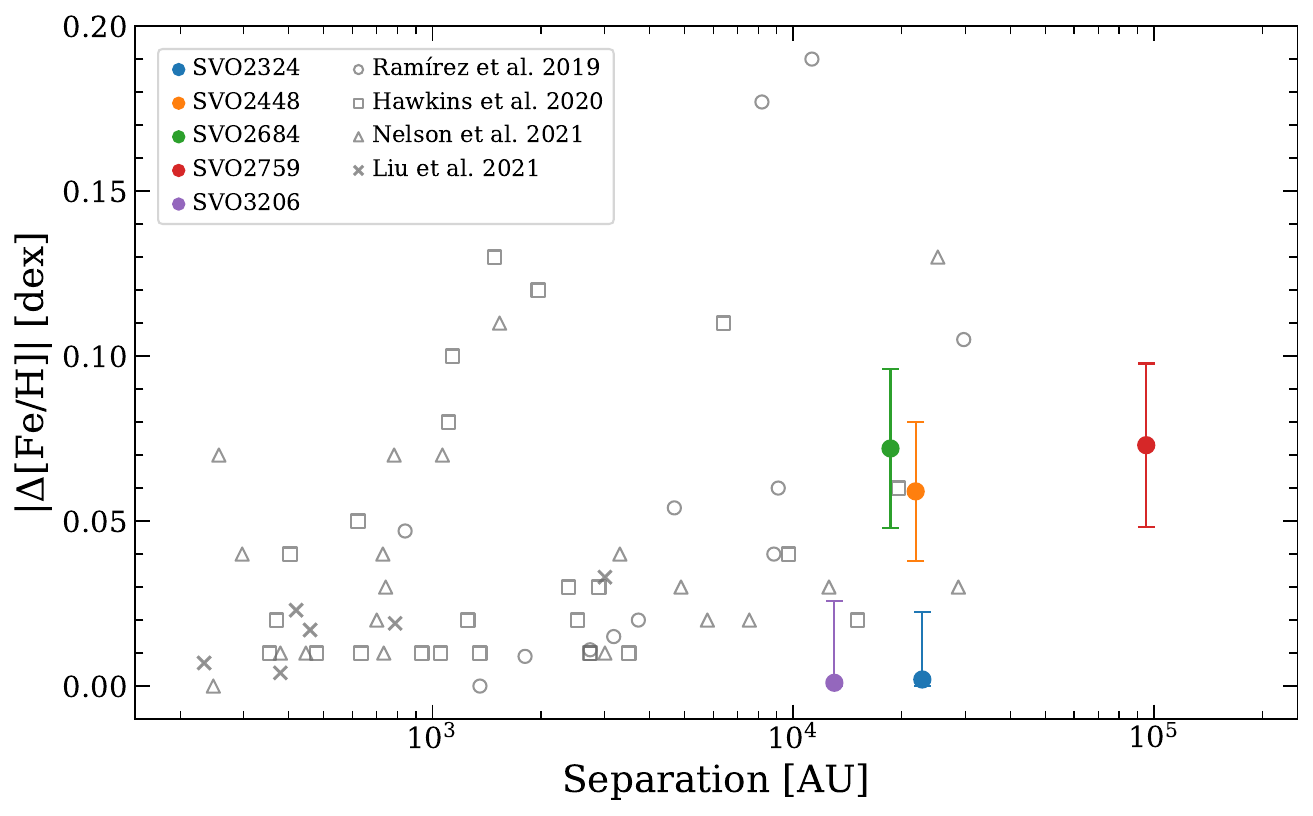} 
     \caption{
     Absolute differences in [Fe/H] as a function of separation between wide binary component stars. 
     Our samples are indicated in filled circles with vertical $\pm$1$\sigma$ error bars estimated as the squared sum of [Fe/H] measurement error for each component star. 
     Open symbols represent the data obtained from \citet{Ramirez2019}, \citet{Hawkins2020}, \citet{Nelson2021}, and \citet{Liu2021}.
     }
     \label{fig:feh_sep}
\end{figure}

Since our samples were selected from wide binary candidates with separations larger than 10,000~AU (see Section~\ref{sec:sub:target}), they can provide significant constraints for the trend of chemical differences with binary separation.
In Figure~\ref{fig:feh_sep}, we present the absolute difference in [Fe/H] as a function of projected binary separation for our samples and data from the literature. 
Notably, the SVO2759 (represented by the red circle) exhibits a moderate difference in [Fe/H] at the largest separation ($\sim$100,000~AU).
The correlation between the $|\Delta$[Fe/H]$|$ and separation is not clearly evident from this figure.
While there appears to be a trend of increasing variation in $|\Delta$[Fe/H]$|$ for wide binaries with larger separations, it is essential to note that there are samples with small differences in [Fe/H] even at large separations.
In particular, two of our sample pairs, SVO2324 and SVO3206, exhibit remarkably tiny differences in [Fe/H] ($<$ 0.01~dex) compared to other wide binaries with separations larger than 10,000~AU. 
This finding suggests that the difference in Fe abundance between the component stars may be influenced by the specific properties of each wide binary pair, which could arise from different formation mechanisms.
For instance, wide binary systems showing smaller differences in [Fe/H] between their component stars, such as SVO2324 and SVO3206, may have formed closer together and then drifted apart through, for example, dynamical unfolding from a higher-order system \citep{Elliott2016}.
Our result highlights the importance of individual characteristics of wide binary pairs based on the assumption that the formation and evolution mechanisms can vary depending on the individual wide binary system \citep[see also][]{Oh2018, Lim2021}. 
On the other hand, it is necessary to be cautious when combining data from multiple sources, as the differences in precision according to data should be taken into account.

\begin{figure}[t]
\centering
   \includegraphics[width=0.48\textwidth]{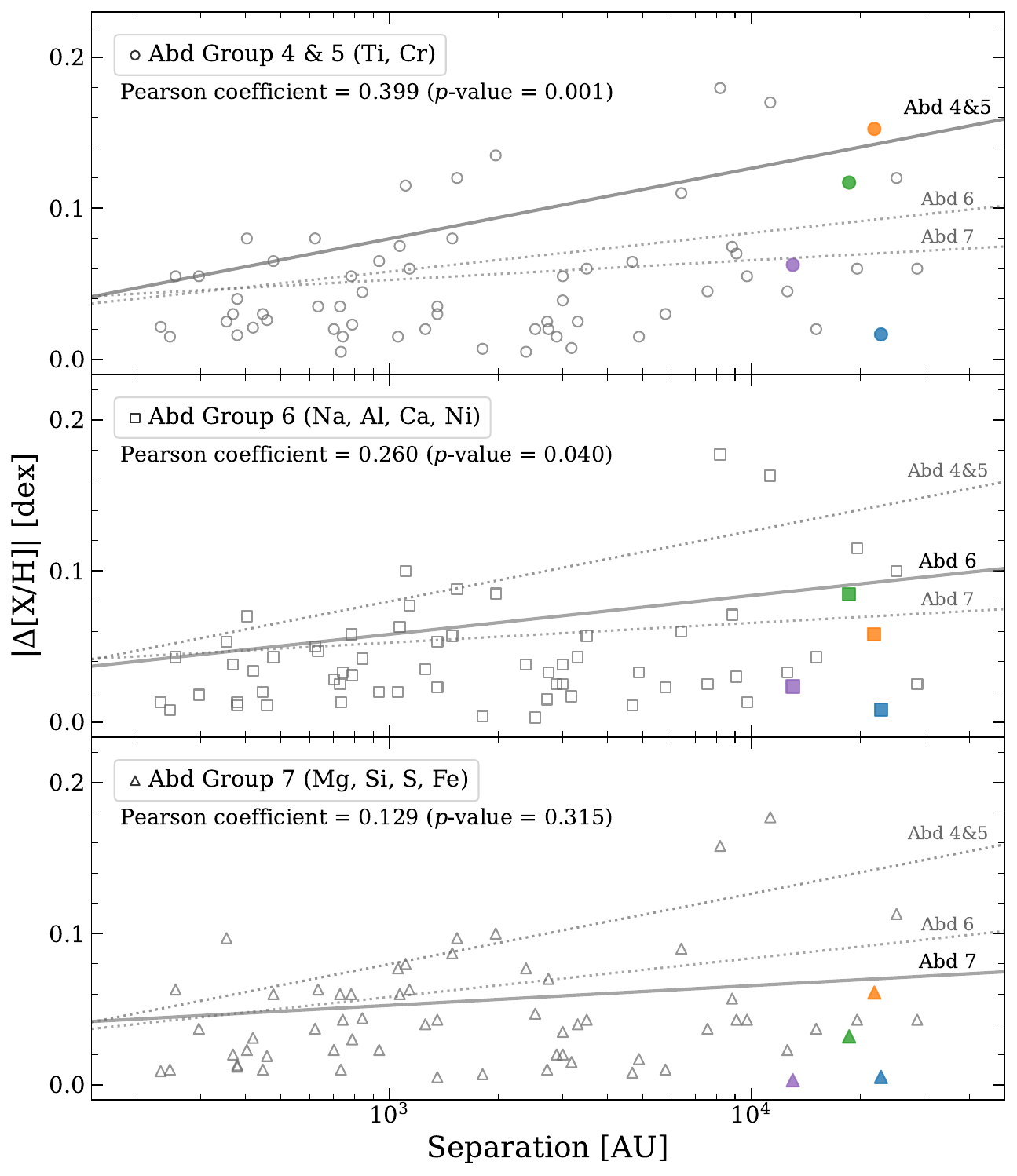} 
     \caption{
     Average absolute differences in three abundance groups, which are classified according to the solar abundance. 
     Four of our samples are indicated in colored symbols, except SVO2759, and open symbols are literature data listed in Figure~\ref{fig:feh_sep}.
     Solid and dashed lines are the least square-fitted lines for each group.
     These lines appear to be located above the data points because the x-axis is plotted on a log scale. 
     }
     \label{fig:abund_sep}
\end{figure}

Furthermore, \citet{Ramirez2019} reported increasing trends in abundance differences between wide binary pair stars with increasing separation, not only for Fe but also for other elements. 
These trends are more pronounced among elements with lower absolute abundance in the Sun, indicating that these elements were less homogeneous in the gas cloud where the wide binary formed.  
\citet{Ramirez2019} categorized elements into seven groups based on their absolute abundance in the Sun, for example, with elements having solar abundance from 2 and 3 are classified as `Abd group 2' (see their Table~7).
In this study, we also classified the elements we measured according to these criteria and plotted the average abundance differences as a function of binary separation in Figure~\ref{fig:abund_sep}. 
However, we limited our analysis to three abundance groups, Abd 4 \& 5 (Ti, Cr), Abd 6 (Na, Al, Ca, Ni), and Abd 7 (Mg, Si, S, Fe), due to a shortage of measured elements.

Figure~\ref{fig:abund_sep} shows trends of increasing variation in absolute differences for the three groups with larger separations, although clear correlations are not evident. 
These trends are comparable to the observed trend in Fe abundance (see Figure~\ref{fig:feh_sep}).  
In addition, the least square-fitted lines for each group reveal that elements with lower solar abundances (Abd 4 \& 5) exhibit a stronger correlation compared to the other groups (Abd 6 and Abd 7).
We performed a Pearson correlation test on wide binaries with a separation of less than 50,000~AU, as farther samples could significantly affect the test. 
The Pearson correlation coefficients and their $p$-values, indicated in each panel of Figure~\ref{fig:abund_sep}, also support weaker correlations in the group of higher solar abundance elements.
Our results remain consistent even when excluding two samples with larger differences in chemical abundances at around $10^{4}$~AU of separation.
These findings suggest that the variation of trends with separation, depending on chemical elements, is a general characteristic of wide binary system, corroborating the claims made by \citet{Ramirez2019}.
However, it is essential to note that the effect of abundance measurement precision among element groups should not be disregarded, as the average abundance difference for Abd group 4 \& 5 is derived from two elements, while the other groups include four elements. 

\subsection{Trend with condensation temperature}\label{sec:sub:trend_2}
An intriguing aspect that can be explored with wide binary systems is the investigation of their star-to-planet interactions.
Since the component stars of wide binaries possess similar chemical properties and are far enough apart not to influence each other, the observed variations in abundance differences depending on the condensation temperature of elements can provide valuable insights into the properties of any potential hosting planet(s).
In this regard, numerous studies have been conducted to examine the existence, formation, and chemical composition of planets or planetary engulfment events \citep[e.g.,][]{Oh2018, Ramirez2019, Jofre2021, Liu2021, Ryabchikova2022}.
A key component of these investigations is comparing the abundance differences between volatile elements (with low condensation temperature, e.g., C, N, O) and refractory elements (with high condensation temperature, e.g., Fe, Ca, Mg), which, in turn, compose gaseous and terrestrial planets.
Wide binary systems, affected by gaseous or terrestrial planets, exhibit noticeable trends in such comparisons.

\begin{figure}
\centering
   \includegraphics[width=0.48\textwidth]{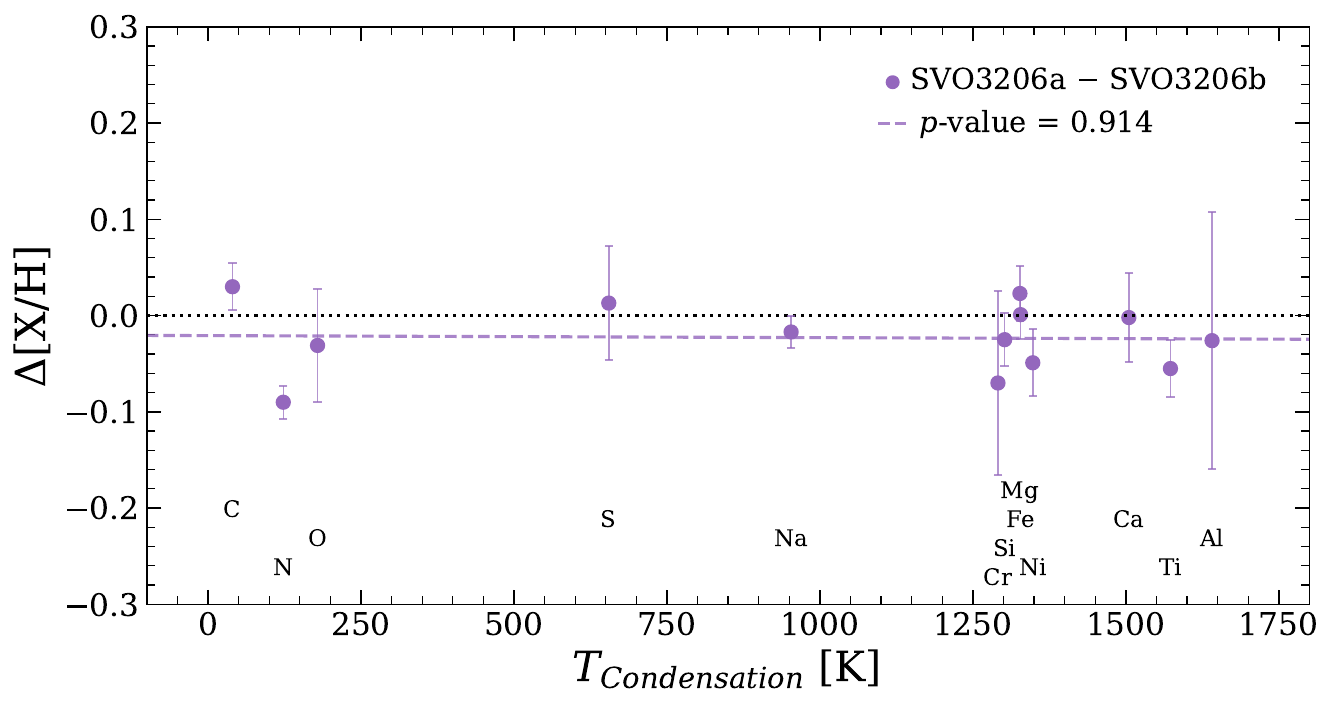} 
     \caption{
     Difference of [X/H] between SVO3206a and SVO3206b stars, with $\pm$1$\sigma$ error, as a function of condensation temperature of elements. 
     The dashed line indicates a least-square fit, and the dotted line denotes the value of 0.0~dex in $\Delta$[X/H].  
     The $p$-value of the Pearson correlation test is written in the upper right corner.
     }
     \label{fig:t_cond}
\end{figure}

In this study, we were able to measure the chemical abundances of C, N, and O elements only for SVO3206 pair stars due to the high ${\rm T_{eff}}$ of other samples. 
Figure~\ref{fig:t_cond} displays $\Delta$[X/H] for SVO3206 (SVO3206a $-$ SVO3206b) as a function of the condensation temperature of elements obtained from \citet{Lodders2003}. 
This plot reveals the absence of any correlation within this wide binary system, as confirmed by a large $p$-value of the Pearson correlation test ($p$-value = 0.914 with coefficient = $-$0.033).
Our result suggests that these two stars have similar planetary system scales, including cases without planets, or both stars are not significantly affected by star-to-planet interactions. 
It is worth noting that no planets have been reported for either star thus far. 
In addition, apart from the lack of correlation with condensation temperature, SVO3206b is consistently more enhanced in chemical abundances than SVO3206a by an average of $\sim$0.02~dex, which is a minor value compared to the other wide binary systems. 

Although we found no trend in chemical abundance differences corresponding to the condensation temperature of elements, our result demonstrates the capability of NIR spectroscopy for planet-hosting stars. 
In particular, the NIR region contains numerous CN, CH, and OH molecular features, which enable more accurate chemical abundance measurements for C, N, and O elements, compared to optical spectroscopy that relies on only a few spectral lines. 
The advantage of NIR spectroscopy for cool stars is also encouraging, given that many exoplanet studies focus on FGK$-$type stars. 
Therefore, the utilization of NIR high-resolution spectroscopy, combined with optical data, will be a powerful tool for studying wide binary systems hosting exoplanets based on a plenty number of chemical element abundances with high-precision. 

\section{Summary and conclusion}\label{sec:concl}
In this study, we conducted high-resolution NIR spectroscopic observations of six common proper motion pairs using the IGRINS spectrograph at the Gemini-South telescope.
Based on the derived RV and chemical abundances for each star, five pairs were confirmed as coeval wide binary systems, which show similar properties between component stars.
In addition, while no clear correlation was found between abundance differences and binary separation, the variation in these differences tended to increase in wide binaries with larger separations. 
Interestingly, two of our wide binary samples, SVO2324 and SVO3206, demonstrated minimal differences in most elements despite their large binary separation.
This result supports the idea of multiple formation mechanisms depending on each wide binary system, emphasizing the need for considering their chemical and dynamical properties.

Furthermore, our study demonstrates the validity of high-resolution NIR spectroscopy for research on wide binary systems, as it accurately observed similar RVs and chemical compositions between the component stars with a high level of precision.
This result offers the potential to examine more wide binary samples, especially those with low temperatures or those located in the Galactic bulge.
Moreover, detailed chemical abundances of volatile elements, such as C, N, and O, provide valuable opportunities for studying exoplanets and their host stars. 
Given the increasing number of exoplanet discoveries and the fact that around 100 planetary systems have been identified within wide binary systems to date, NIR spectroscopy will play a more prominent role in the study of exoplanets and their host stars.
However, to further enhance these capabilities, it remains crucial to extend the measurable spectral features and to augment the number of elements with qualified information. 

We will continue high-resolution spectroscopic observations for more wide binaries in the Milky Way, focusing on general samples as well as those with peculiar chemical properties. 
This includes examining depleted $\alpha$-element abundances of accreted objects and discrepancies between volatile and refractory element abundances of planet-host stars. 
These observations will contribute to a better understanding of various formation processes of wide binary systems and provide valuable insights into many aspects of Galactic astronomy.


\begin{acknowledgments}
We thank the referee for a number of helpful suggestions.
DL, SH, and YWL acknowledge support from the National Research Foundation of Korea to the Center for Galaxy Evolution Research (2022R1A6A1A03053472 and 2022R1A2C3002992). 
S.H.C. acknowledges support from the National Research Foundation of Korea (NRF) grant funded by the Korea government (MSIT) (NRF-2021R1C1C2003511) and the Korea Astronomy and Space Science Institute under R\&D program (Project No. 2023-1-830-00) supervised by the Ministry of Science and ICT.
DL thanks Sree Oh for the consistent support. 
This work was supported by K-GMT Science Program (PID: GS-2021B-Q-310 and GS-2023A-Q-309) of Korea Astronomy and Space Science Institute (KASI).
This work used the Immersion Grating Infrared Spectrometer (IGRINS) that was developed under a collaboration between the University of Texas at Austin and the KASI with the financial support of the Mt. Cuba Astronomical Foundation, of the US National Science Foundation under grants AST-1229522 and AST-1702267, of the McDonald Observatory of the University of Texas at Austin, of the Korean GMT Project of KASI, and Gemini Observatory.
\end{acknowledgments}

\vspace{5mm}
\facilities{Gemini:South (IGRINS)}

\software{IGRINS Pipeline Package \citep{Lee2017}, 
	astropy \citep{AstropyCollaboration2013, AstropyCollaboration2018}, 
	MOOG \citep{Sneden1973} 
	}

\bibliography{export-bibtex}{}
\bibliographystyle{aasjournal}

\end{document}